\documentclass{aa}
\usepackage{graphics}
\def\msun{\ifmmode {\rm M_\odot} \else M$_\odot$\fi}
\def\msunyr{\ifmmode {\rm M_\odot~yr^{-1}}\else${\rm M_\odot~yr^{-1}}$\fi}

\def\gtsim{\raisebox{-.5ex}{$\;\stackrel{>}{\sim}\;$}}

\newcommand{\ha}{\mbox{H$\alpha$}}
\newcommand{\hb}{\mbox{H$\beta$}}

\newcommand{\bra}{\mbox{Br$\alpha$}}

\newcommand{\oiii}{\mbox{[O\,{\sc iii}]}}
\newcommand{\feii}{\mbox{Fe\,{\sc ii}}}

\newcommand{\funits}[1]
 {\mbox{10$^{-#1}\,$ergs\ s$^{-1}$\,cm$^{-2}$\,\AA$^{-1}$}}
\newcommand{\lunits}[1]{\mbox{10$^{-#1}\,$ergs\ s$^{-1}$\,cm$^{-2}$}}

\newcommand{\kms}{\mbox{km\,s$^{-1}$}}

\newcommand{\mic}{\mbox{$\mu$m}}

\begin{document}

     \thesaurus{03               
               (11.01.2;         
               11.09.1: Mrk 279; 
               11.14.1;          
               11.19.1;          
               11.06.2: molecular torus;  
               13.09.1)}         

 \title{
   Monitoring of the optical and 2.5 -- 11.7 $\mu$m spectrum and
   mid-IR imaging of the Seyfert 1 galaxy Mrk~279 with ISO
   \thanks{Based on observations collected with
   the Infrared Space Observatory, ISO, an ESA project with
   instruments  funded by ESA member states
   (especially the PI countries: France,
   Germany, The Netherlands, and the United Kingdom) with the participation
   of ISAS and NASA.}
  }
\author{
  M.~Santos-Lle\'o\inst{1} \and
  J.~Clavel\inst{1} \and
  B.~Schulz,\inst{2}
  B.~Altieri,\inst{1}
  P.~Barr\inst{3} \and
  D.~Alloin\inst{4} \and
  P.~Berlind\inst{5} \and
  R.~Bertram\inst{6}$^,$\inst{7}
  D.M.~Crenshaw\inst{8} \and
  R.A.~Edelson\inst{9} \and
  U.~Giveon\inst{10} \and
  K.~Horne\inst{11} \and
  J.P.~Huchra\inst{5} \and
  S.~Kaspi\inst{10} \and
  G.A.~Kriss\inst{12} \and
  J.H.~Krolik,\inst{13}
  M.A.~Malkan\inst{14} \and
  Yu.F.~Malkov\inst{15}
  H.~Netzer\inst{10} \and
  P.T.~O'Brien\inst{9} \and
  B.M.~Peterson\inst{6} \and
  R.W.~Pogge\inst{6} \and
  V.I.~Pronik\inst{15}$^,$\inst{16} \and
  B.-C.~Qian\inst{17} \and
  G.A.~Reichert\inst{18} \and
  P.M.~Rodr\'{\i}guez-Pascual\inst{19} \and
  S.G.~Sergeev\inst{15}$^,$\inst{16} \and
  J.~Tao\inst{17} \and
  S.~Tokarz\inst{5} \and
  R.M.~Wagner\inst{6}$^,$\inst{7} \and
  W.~Wamsteker\inst{20} \and
  B.J.~Wilkes\inst{5}
  }

  \offprints{M. Santos-Lle\'o}
  \mail{msantos@xmm.vilspa.esa.es}

 \institute{XMM Science Operations Center, Astrophysics Division, ESA Space
    Science Department, P.O. Box  50727, E-28080 Madrid, Spain
  \and ISO Science Operations Center, Astrophysics Division, ESA Space
    Science Department, P.O. Box  50727, E-28080 Madrid, Spain
  \and Integral Science Operations Center, Astrophysics Division, ESA
    Space Science Department, ESTEC, Postbus 299, 2200 AG Noordwijk, The
    Netherlands
  \and European Southern Observatory, Alonso de Cordova 3107,
    Vitacura Casilla 19001, Santiago 19, Chile
  \and Harvard-Smithsonian Center for Astrophysics, 60 Garden Street,
    Cambridge, MA  02138, USA
  \and Department of Astronomy, Ohio State University, 140 West 18th Avenue,
    Columbus, OH 43210, USA
  \and Mailing address: Steward Observatory, University of Arizona,
    Tucson, AZ  85721, USA
  \and Computer Sciences Corporation, Laboratory for Astronomy and Solar
    Physics, NASA Goddard Space Flight Center, Code 681, Greenbelt, MD 20771,
    USA
  \and Department of Physics and Astronomy, University
    of Leicester, University Road, Leicester LE1 7RH, UK
  \and School of Physics and Astronomy and the Wise Observatory, The Raymond
    and Beverly Sackler Faculty of Exact Sciences, Tel-Aviv University,
    Tel-Aviv 69978, Israel
  \and School of Physics and Astronomy, University of St. Andrews, North
    Haugh, St. Andrews KY16 9SS, Scotland, UK
  \and Space Telescope Science Institute, 3700 San Martin Drive,
    Baltimore, MD 21218, USA
  \and Department of Physics and Astronomy, The John Hopkins University,
    Baltimore, MD 21218, USA
  \and Department of Astronomy, University of California, Math-Science
    Building,  Los Angeles, CA 90024, USA
  \and Crimean Astrophysical Observatory, P/O Nauchny, 334413 Crimea,
        Ukraine
  \and Isaac Newton institute of Chile, Crimean Branch
  \and Shanghai Astronomical Observatory, 80 Nandan Road, 200030 Shanghai,
        People's Republic of China
  \and Raytheon ITSS, Space Science Data Operations Office,
    NASA Goddard Space Flight Center, Code 631, Greenbelt MD 20771, USA
  \and Universidad Europea de Madrid, Departamento de F\'\i sica, C/ Tajo
    sn, Urb. El Bosque, Villaviciosa de Od\'on, 28670 Madrid, Spain
  \and ESA IUE Observatory, P.O. Box 50727, 28080 Madrid, Spain
  }


  \date{Received dd mmm yyyy / Accepted dd mmm yyyy }

  \titlerunning{Monitoring of the mid-infrared flux of Mrk~279 with ISO}
  \authorrunning{M. Santos Lle\'o et al. }

\maketitle

\begin{abstract}

Mid-infrared images of the Seyfert~1 galaxy \object{Mrk~279}
obtained with the ISO satellite are presented together with the
results of a one-year monitoring campaign of the 2.5--11.7\,\mic\ spectrum.
Contemporaneous optical photometric and spectrophotometric observations
are also presented.
The galaxy appears as a point-like source at the resolution of the
ISOCAM instrument (4--5\,$\arcsec$).
The 2.5--11.7\,\mic\ average spectrum of the nucleus in Mrk~279 shows a
strong power law continuum with $\alpha\,=\,-0.80\pm0.05$ (${\rm
F_{\nu}\,\propto\,\nu^{\alpha}}$) and weak emission PAH features.
The Mrk~279 spectral energy distribution shows a mid-IR bump,
which extends from 2 to 15--20\,\mic . The mid-IR bump is
consistent with thermal emission from dust grains at a distance of $\gtsim
100$\,lt-d.
No significant variations of the mid-IR flux have been detected during our
observing campaign, consistent with the relatively low amplitude ($\sim$ 10\,\%
rms) of the optical variability
during the campaign. The time delay for \hb\  line emission in response to the optical
continuum variations is $\tau = 16.7^{+5.3}_{-5.6}$\,days, consistent with
previous measurements.

\keywords{galaxies: active -- galaxies: individual:
Mrk\,279 -- galaxies: nuclei -- galaxies: Seyfert
-- galaxies: molecular torus -- Infrared: galaxies}

\end{abstract}

\section{Introduction}

According to the unified model of active galactic nuclei (AGN), the central
massive black hole, its surrounding accreting material and the
broad-line region (BLR) are all embedded within a dusty region,
probably a thick molecular torus.
Along some directions, the dust extinction is
sufficient to block all UV, optical, and
near-IR radiation originating in
the inner components.
A review of the arguments which led to this picture is presented
in Wills (1999).
The presence of a universal inflection point near 1.2\,\mic\ 
in the spectral energy distribution of radio-quiet AGN's strongly 
suggests that the bulk of the near IR flux arises from dust thermal emission
(e.g., Barvainis 1987; Sanders et al. 1989). The corresponding 
color temperature, $\simeq$ 1,500\,K, matches closely the sublimation 
temperature of graphite, the most resilient of the grain constituents.
The near IR emission can be variable, and therefore originates, at 
least in part, in a compact region. Furthermore, in three AGN,
the near IR variations have been shown to be
delayed with respect to the
UV-optical variations. The time delay
corresponds closely to the light-travel time to the
dust sublimation radius $r_{\rm in}$; measured values
of $r_{\rm in}$ are
400 light days for \object{Fairall~9} (Clavel et al.\ 1989),
50 light days for \object{NGC~1566} (Baribaud et al.\ 1992),
80 light days for \object{NGC~3783} (Glass 1992) and 32 light days 
for Mrk 744 (Nelson, 1996).

The emerging picture is one where the near to mid-IR
emission arises from thermal re-radiation of UV and optical photons
absorbed by the circumnuclear dust.
Various models for the geometry and location of this dust
have been proposed, but the exact configuration
of the models remains unconstrained due to a lack of suitable
observational data.

One can use variability as a tool 
to probe the {\em inner}\ few light\,years of the dusty regions.
Reverberation-mapping techniques (Blandford \& McKee 1982)
have been used extensively to map the BLR in several AGN,
on scales of light\,days to light\,months, notably by
the International AGN Watch\footnote{For a complete panorama
of the AGN Watch data sets, results, and
related studies, see the AGN Watch web page at URL
{\sf http://www.astronomy.ohio-state.edu/$\sim$agnwatch/}.}
consortium (Alloin et al.\ 1994).
A similar approach can be used to probe the IR-emitting region,
i.e. the warm dust component within the obscuring material.
Given UV flux variations of sufficient amplitudes, a mid-IR
monitoring campaign of sufficiently long duration and adequate
sampling rate, it may in principle be possible to recover the transfer 
function of the dust.

The Infrared Space Observatory (ISO;
Kessler et al.\ 1996) offerred a unique opportunity to  carry out
such a spectrophotometric monitoring program in the mid-IR.
The Seyfert 1 galaxy Mrk~279
($z = 0.0294$) was selected
because its celestial position allows an uninterrupted 12-month visibility
window for ISO and it has a well-documented variability history in the
optical (Osterbrock \& Shuder 1982; Peterson et al.\ 1985;
Maoz et al.\ 1990; Stirpe et al.\ 1994, the UV (Chapman et al.\ 1985), and
X-rays (Reichert et al.\ 1985).
Balmer-line time-delays (Maoz et al. 1990; Stirpe \& de Bruyn 1991;
Stirpe et al. 1994)
suggest a BLR size in the range 6 to 12 light\,days.
A search for day-to-day variability across the Balmer-line profile
was unsuccessful (Eracleous \& Halpern 1993). No far-IR flux variations were
detected with IRAS (Edelson \& Malkan 1987).

\section{IR observations and data reduction}

Mrk~279 was observed with two of the instruments on board the
ISO satellite: nine narrow-band filter images were obtained with
ISOCAM (Cesarsky et al.\ 1996),
while spectra were recorded with the
PHT-S spectrometer, a subsystem of the ISOPHOT
instrument (Lemke et al.\ 1996).
The PHT-S spectrometer covers the 2.5--12~\mic\ spectral range at a mean 
resolution of $\sim$3150\,\kms, with a gap between 4.9 and 5.9\,\mic. 
Its entrance aperture projects to $24\arcsec \times 24\arcsec$ on
the sky.
All PHT-S observations were carried-out in an identical fashion. The
integrating amplifiers were reset every 32\,s and on-source
measurements were interleaved with background measurements by ``chopping''
on the sky at a frequency of 1/256 Hz. The chopper throw was set to 300$''$.
For each of the observations, total on-source integration time was 2048~s
and total observing time (including background measurements and instrument
overheads) was 4236~s.

The CAM observations were performed in staring mode at a magnification
of $3\,\arcsec$ per pixel and with a gain of 2. Nine different filters were
used. The unit integration time was 2.1\,s per readout and there were between 
72 and 197 readouts per image, depending on the filter used. The particular 
sequence of filters was chosen to go from high to low illumination so as to 
minimize the detector stabilization time. Good stabilization was also
guaranteed by the relatively large number of readouts per exposure.

The PHT-S observations were made at 16 different epochs, from 1996, February
5 to 1997, February 13 (Table~1). The ISOCAM observations of Mrk~279 were
all carried out on 1996 February 5, contemporaneously with
the first of the PHT-S spectra.

Standard procedures from the CAM Interactive Analysis (CIA) software package
\footnote{CIA is a joint development by the ESA Astrophysics Division and
the CAM Consortium.} were used for the reduction of the ISOCAM data
(Ott et al.\ 1997). The full width at half maximum (FWHM) of Mrk~279 varies 
with the filter wavelength from $3''\!.3$ to $5''\!.0$, (table~\ref{filters})
but is always consistent with that of a point source.
Monochromatic intensities were obtained by integration of the source flux
within a circle of radius 6 pixels ($18''$) and subtraction of a 
normalised background measured in a concentric circular ring.
The intensity of Mrk~279 in the different filters is listed in 
Table~2. The accuracy of these measurements is $\pm 10$\,\%.

The ISOPHOT-S data were reduced with the PHOT Interactive Analysis (PIA;
  \footnote{PIA is a joint development by the ESA Astrophysics Division and 
  the ISOPHOT Consortium led by the ISOPHOT PI, D. Lemke, MPIA, Heidelberg.} 
  Gabriel \cite{gabriel}) software package. However, because  ISOPHOT-S was 
  operating close to its sensitivity limit, special reduction and 
  calibration procedures had to be applied. After a change of illumination, 
  the responsivity of the Si:Ga photoconductors immediately jumps to an 
  intermediate level. This initial jump is followed by a characteristic slow 
  transient to the final level. At the faint flux limit, this time constant is 
  extremely long, and in practice only the initial step is observed in 
  chopped-mode. The spectral response function for this particular mode and 
  flux-level was derived directly from observations of a faint 
  standard star HD~132142 whose flux ranges from 0.15 to 2.54 Jy. The 
  calibration star observation was performed with the same chopper frequency 
  and readout-timing as the AGN observations. The $S/N$ of the ISOPHOT-S 
  spectra was considerably enhanced by two additional measures: 
  $i$) the 32-s integration ramps  were divided into sub-ramps of 2~sec 
  and no de-glitching (removal of cosmic ray hits) was performed at 
  ramp-level $ii$) after slope-fitting and de-glitching at slope-level, 
  the maximum of the distribution of the slopes was determined by 
  fitting a gaussian to the histogram. The resulting ISOPHOT-S fluxes are
  accurate to within $\pm10$~\%.

\begin{figure}
      \resizebox{0.82\hsize}{!}{\includegraphics{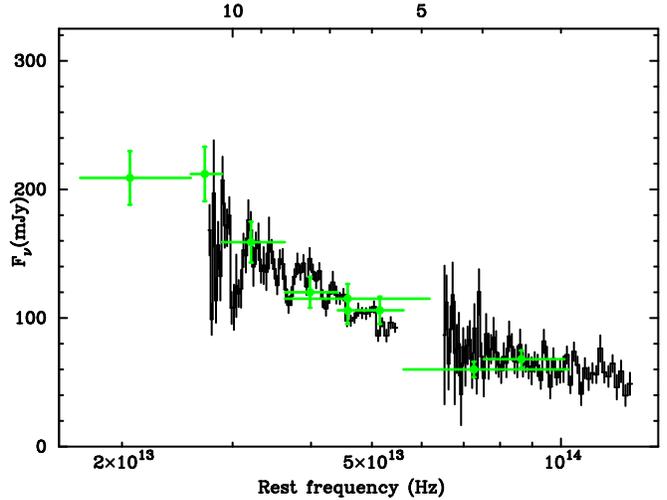}}
      \caption{The ISOPHOT-S spectrum from 1996 February 5 together
       with the ISOCAM photometric fluxes from the same date. The
       x-axis shows the rest frequency at the bottom and the rest
       wavelength in micron at the top, both on a logarithmic scale}
      \label{feb05}

   \end{figure}

The 1996 February 5 PHT-S spectrum is shown in Fig.~\ref{feb05}, together
with the monochromatic intensities measured with ISOCAM on the same day. 
Each PHT-S data-point is shown with its formal error as propagated by the 
PIA software.
The CAM and PHT-S fluxes agree to better than $\pm10$\,\%, providing
further confidence in the reliability of the flux calibration. 

\section{MIR flux Measurements and uncertainties}

\subsection {PHT-S flux measurements and reproducibility}

\label{irmeasure}

For each epoch of observation the continuum flux
was measured over two different intervals, at short wavelengths (SW: 
2.5--4.7\,\mic) and long wavelengths (LW: 5.8--9.9\,\mic). 
Table~\ref{lcur}  lists the mean intensities
over these intervals and their uncertainties, while the light curves are
shown in Fig.~\ref{lcplot}. 

An accurate determination of the flux uncertainties is essential when 
discussing source variability. We have therefore investigated the different
source of errors which could potentially affect our PHT-S measurement.

\label{stabil}

In staring mode, the overall responsivity of PHT-S is known to remain 
stable within $\pm\,10$\% (Schulz 1999).
We have assessed the stability of the PHT-S responsivity more
specifically at the time of each of the Mrk~279 observations and verified
that no other systematic effects were present. For this purpose,
two different types of calibration measurements were used as diagnostic:
\begin{enumerate}
\item {The detector dark current measurements, which are obtained immediately
prior to each PHT-S observation: throughout the campaign, the dark current 
signal retained its nominal value of $\sim$ zero $V/s$. We can therefore
be confident that none of the Mrk~279 observations suffered from detector
remanence induced by a prior exposure to a bright source.}
\item {Measurements with an internal calibration source which are carried-out
systematically at the beginning of each ISO revolution: averaged over all 
pixels, the variations of detector responsivity from epoch to epoch are 
$\simeq\,\pm\,2$\%, with upper limits of 3\% and 5\% in the SW and LW 
range, respectively.}
\end{enumerate}
A conservative upper limit of $\pm10$\% was thus adopted for the 
{\em systematic\/} uncertainty on the PHT-S fluxes of Mrk~279. 
Internal measurement errors were added in quadrature to this systematic
uncertainty. The internal errors
were computed as the dispersion about the mean flux in the SW
and LW integration intervals, after normalization of
the spectra. The purpose of the normalization is
to remove the spectral curvature.
Each spectrum was first divided by its best-fit power-law continuum
($ F_\nu = 8.44\, 10^{-14} \nu ^{-0.8}$ erg\,s$^{-1}$cm$^{-2}$Hz$^{-1}$;
see \S{\ref{irspe}}) and the rms deviations were computed. The results 
show that the internal errors associated
with each flux measurement are of 2.8\,\% and 1.7\,\% for the
2.52-4.70\,\mic\ and 5.76-9.89\,\mic\ bands, respectively.

As a consistency check, errors were also computed by comparing PHT-S fluxes 
obtained within 30 days from each others. This gives a conservative error 
estimate since it assumes that there are no flux variations on time scales
shorter than 30 days. Taking every pair of fluxes within 30 days and
measuring the error on their means, we get mean relative errors of
3.5\,\% and 1.6\,\% for the SW and LW bands, respectively.

\begin{center}
\begin{table}[htbp]
\caption{PHT-S observation log and fluxes}
\label{lcur}
\begin{tabular}{lccc}
\multicolumn{1}{c}{UT} &
\multicolumn{1}{c}{MJD} &
\multicolumn{1}{c}{$F$(2.5--4.7\,$\mu$)} &
\multicolumn{1}{c}{$F$(5.8--9.9\,$\mu$)} \\
\multicolumn{1}{c}{} &
\multicolumn{1}{c}{(-2,450,000)} &
\multicolumn{1}{c}{(mJy)}&
\multicolumn{1}{c}{(mJy)} \\
\multicolumn{1}{c}{(1)} &
\multicolumn{1}{c}{(2)} &
\multicolumn{1}{c}{(3)} &
\multicolumn{1}{c}{(4)} \\
1996 Feb    5  &  119  & 73.5  $\pm$ 7.6 & 132 $\pm$ 13  \\
1996 Mar    3  &  146  & 75.3  $\pm$ 7.8 & 123 $\pm$ 12  \\
1996 Mar   12  &  155  & 65.2  $\pm$ 6.8 & 125 $\pm$ 13  \\
1996 Apr    2  &  176  & 69.5  $\pm$ 7.2 & 129 $\pm$ 13  \\
1996 Apr   27  &  201  & 72.7  $\pm$ 7.6 & 127 $\pm$ 13  \\
1996 May   11  &  215  & 72.8  $\pm$ 7.6 & 128 $\pm$ 13  \\
1996 May   29  &  233  & 73.2  $\pm$ 7.6 & 131 $\pm$ 13  \\
1996 Jul   29  &  294  & 63.6  $\pm$ 6.6 & 127 $\pm$ 13  \\
1996 Aug   12  &  308  & 61.8  $\pm$ 6.4 & 122 $\pm$ 12  \\
1996 Aug   27  &  323  & 62.0  $\pm$ 6.4 & 123 $\pm$ 12  \\
1996 Sep   15  &  342  & 60.0  $\pm$ 6.2 & 124 $\pm$ 13  \\
1996 Oct   17  &  374  & 73.8  $\pm$ 7.7 & 128 $\pm$ 13  \\
1996 Nov    1  &  389  & 62.6  $\pm$ 6.5 & 120 $\pm$ 12  \\
1996 Nov   18  &  406  & 73.7  $\pm$ 7.7 & 128 $\pm$ 13  \\
1996 Dec    5  &  423  & 70.1  $\pm$ 7.3 & 124 $\pm$ 13  \\
1997 Feb   13  &  493  & 81.8  $\pm$ 8.5 & 128 $\pm$ 13  \\
\end{tabular}
\end{table}
\end{center}

\begin{center}
\begin{table}[htbp]
\caption{ISOCAM narrow band filter intensities}
\label{filters}
\begin{tabular}{ccccc}
\multicolumn{1}{c}{Filter} &
\multicolumn{1}{c}{$\lambda_c$} &
\multicolumn{1}{c}{Range} &
\multicolumn{1}{c}{Flux} &
\multicolumn{1}{c}{FWHM} \\ 
\multicolumn{1}{c}{} &
\multicolumn{1}{c}{(\mic)} &
\multicolumn{1}{c}{(\mic)} &
\multicolumn{1}{c}{(mJy)}  &
\multicolumn{1}{c}{($''$)} \\
\multicolumn{1}{c}{(1)} &
\multicolumn{1}{c}{(2)} &
\multicolumn{1}{c}{(3)} &
\multicolumn{1}{c}{(4)} &
\multicolumn{1}{c}{(5)} \\
 SW1 & 3.57  &  3.05-4.10 &  68 & 3.9  \\
 SW5 & 4.25  &  3.00-5.5  &  60 & 3.3  \\
 LW4 & 6.00  &  5.50-6.50 & 106 & 4.1  \\
 LW2 & 6.75  &  5.00-8.50 & 115 & 4.8  \\
 LW5 & 6.75  &  6.50-7.00 & 106 & 4.2  \\
 LW6 & 7.75  &  7.00-8.50 & 120 & 5.0  \\
 LW7 & 9.62  &  8.50-10.7 & 159 & 3.8  \\
 LW8 & 11.4  &  10.7-12.0 & 212 & 4.5  \\
 LW3 & 15.0  &  12.0-18   & 209 & 5.0  \\
\end{tabular}
\end{table}
\end{center}

\subsection{Comparison with ground based measurements and estimation of the
host galaxy contribution}
\label{irgal}

Spinoglio et al.\ (1985) measured $L$-band ($\sim3.5\,\mic$) fluxes of
$100\pm21$\,mJy, $112\pm27$\,mJy, and $68\pm15$\,mJy
through apertures of 12$''$, 12$''$, and 17$''$, respectively, consistent 
with our results to within the measurement uncertainties.

Given the spectrograph aperture ($24''\times 24''$),
the host galaxy of Mrk~279 could, in principle, contribute to 
the PHT-S flux. Indeed, a faint extended nebulosity is
apparent in the $K$-band ($\sim 2.2\,\mic$) image of 
McLeod \& Rieke (1995). This extended flux arises from
the integrated emission of giants and supergiants in the galactic 
disk whose energy distribution is maximum at $\sim 2\,\mic$ and 
falls-off abruptly at longer wavelengths. In practice, stellar emission
will therefore make a negligible contribution to the PHT-S 
flux. Nevertheless, this was positively verified by comparison with 
ground-based data as follows:
\begin{enumerate}
\item In the $K$-band, McLeod \& Rieke (1995) estimated that
the AGN contributes 90\% of the flux within the central $1''\!.5$ (FWHM)
and 55\% (35~mJy) of the total $K$-band flux (68~mJy) integrated over the
whole galaxy (i.e., out to a radius of 35$''$). Assuming ``normal'' near-IR
colors ($K-L=0.22\pm0.02$\,mag; Clavel et al.\ 1989) for the stellar
population, we derive a total host-galaxy flux of 18\,mJy at $3.5\,\mic$.
\item Using the $B$-and $R$-band nucleus--galaxy decomposition of Granato
et al.\ (1993), and assuming normal optical-to-IR ($V-K=3.22$\,mag; Clavel 
et al.\ 1989) and $K-L$\, colors, we estimate a total galaxy flux of 
20\,mJy and 17\,mJy respectively at $3.5\,\mic$.
\end{enumerate}
These consistent estimates can be used to infer the amount of 
stellar light which enters the $24''\times 24''$ PHT-S aperture, 
$5\pm2$~mJy at $3.5\,\mic$ flux. Such a small contamination is 
within the measurement errors and can be neglected.

The galaxy contributes 
$11\pm 4$\,mJy, $14\pm 5$\,mJy and $11\pm 4$\,mJy to the $12\,\arcsec$
photometric measurements in the $J$, $H$, and $K$ bands respectively 
(Granato et al. 1993). These values are used in \S{\ref{sedsec}} to infer 
the intrinsic spectral energy distribution of the active nucleus in Mrk~279

\section{Optical observations and data reduction}

Ground-based optical observations were made in support of the
ISO observations. Spectroscopic monitoring was carried out
with the 1.8-m Perkins Telescope of the Ohio State and
Ohio Wesleyan University at the Lowell Observatory,
the 1.0-m telescope of the Wise Observatory,
the 2.6-m Shajn Telescope of the Crimean Observatory,
and the 1.5-m Mt.\ Hopkins Telescope of the Harvard-Smithsonian
Center for Astrophysics (CfA).
A detailed log of the spectroscopic observations can be found at
the International AGN Watch website.

In addition CCD photometry was
made on the 1.0-m telescope of  the Wise Observatory.
The flux was measured using a fixed aperture of radius 7$''$, under seeing
conditions of 2--3$''$. 
The $B$, $V$, $R$, and $I$ instrumental magnitudes are listed in 
columns (3) -- (6) of table~\ref{photom}.
They have not been scaled to any standard system.

The spectroscopic data were processed by the individual observers
in standard fashion for CCD frames. However, the standard astronomical flux
calibration techniques
based on determining the instrument response function
from observations of standard stars are only accurate
for AGN spectrophotometry at
about the 10\% level even under ideal observing conditions.
We used the standard stars only for {\em relative}
calibration, and employed different calibration
techniques for absolute calibration:
The data from Wise Observatory were calibrated following the 
method described by Maoz et al. (1994). The data 
from Ohio, Crimean, and CFA were 
calibrated through
scaling through \oiii\,$\lambda5007$ flux that was measured
on five photometric nights, at F(\oiii\,$\lambda5007$)
= $(1.512\pm0.096) \times $ \lunits{13}.

On these calibrated spectra we measured the continuum flux
by averaging the flux in a 10\,\AA\ wide band centered at 5105\,\AA\ in the
rest frame of Mrk~279 ($F_{\lambda}(5100\,\mbox{\rm \AA})$). The \hb\ line flux
has been measured by linear interpolation between rest-frame wavelengths
$\sim4765$\,\AA\ and $\sim5105$\,\AA\/, and line integration between 4770\,\AA\
and
4935\,\AA. The long-wavelength
cutoff of this integration band misses some of the \hb\ flux
underneath \oiii\,$\lambda4959$, but avoids the need to
estimate the \feii\ contribution to this feature and still gives
a good representation of the \hb\ variability. We did not
correct for the narrow-line, which is expected to be constant.

\begin{figure}[htbp]
      \resizebox{\hsize}{!}{\includegraphics{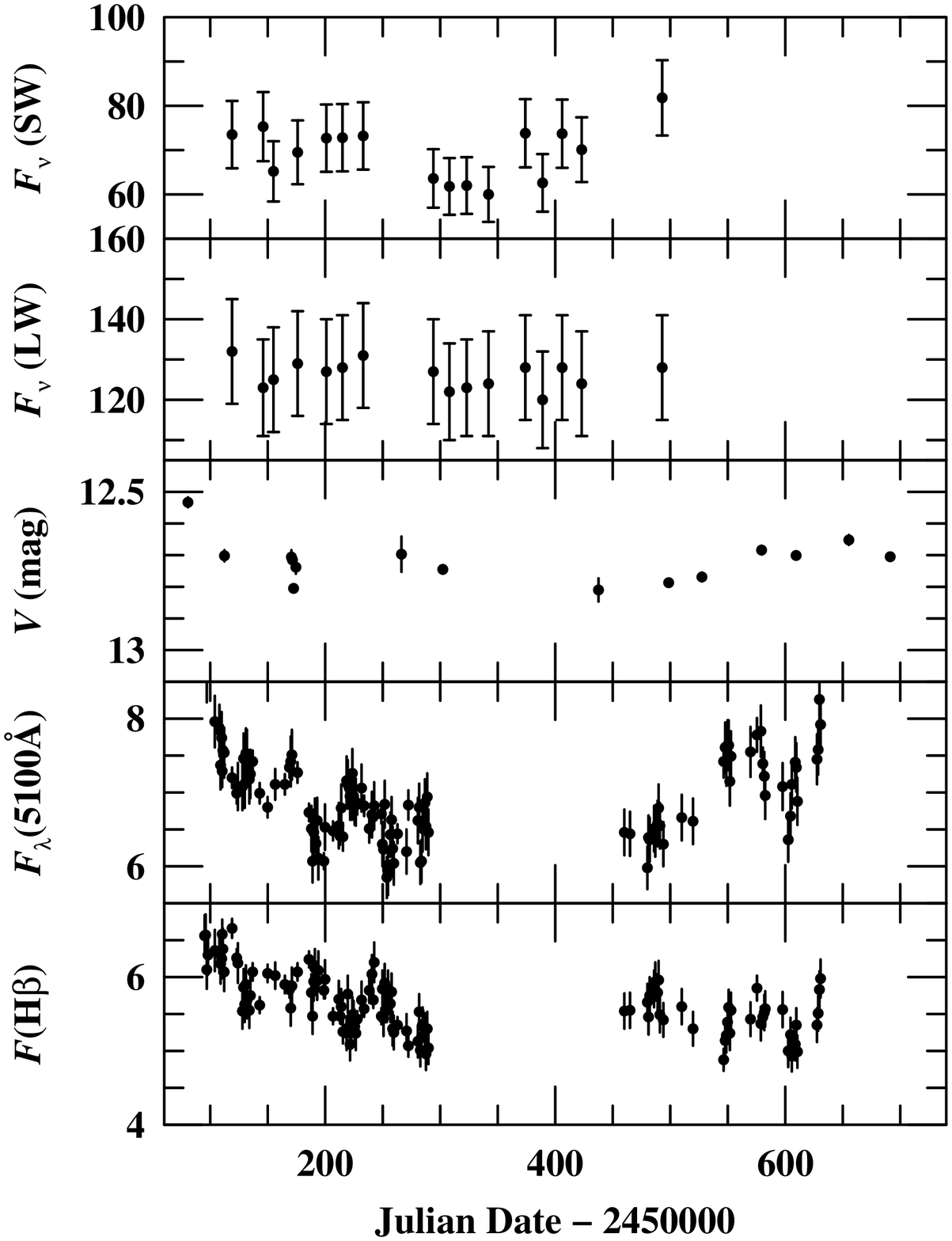}}
      \caption{IR and optical light curves of Mrk~279. The top two panels 
      show the IR light curves from the SW and LW detectors 
      (2.52--4.70 $\mu$m and 5.8--9.9\,\mic , respectively). 
      The third panel shows 
      $V$-band photometric measurements from Wise Observatory.
      The forth panel shows the 5100\,\AA\
      continuum flux measured from the spectra, 
      in units of \funits{15}. The \hb\ flux is shown in the bottom
      panel, in units of \lunits{13}. The optical photometry, optical 
      continuum and \hb\ fluxes are  
      available in electronic form at the CDS.}
      \label{lcplot}
   \end{figure}

As the measurements made from the spectra are subject to systematic 
differences between the four data sets used, we applied the prescriptions of 
Peterson et al.(1999) to intercalibrate the data sets, and correct 
for aperture effects. The final continuum $F_{\lambda}$(5100\,\AA\/) and 
\hb\/ emission-line fluxes are given in table~\ref{optlc}
The spectrophotometric and photometric light curves are
shown in Fig.~\ref{lcplot}. Using the results of Granato et al. (1993), 
we estimate that stars contribute for $1.2\pm$0.6 mJy or $20\pm 10$\% to 
the mean redenning corrected 5100\,\AA\/ flux. 




\section{The mid-IR spectrum of Mrk~279}
\label{irspe}

\begin{figure}
      \resizebox{0.82\hsize}{!}{\includegraphics{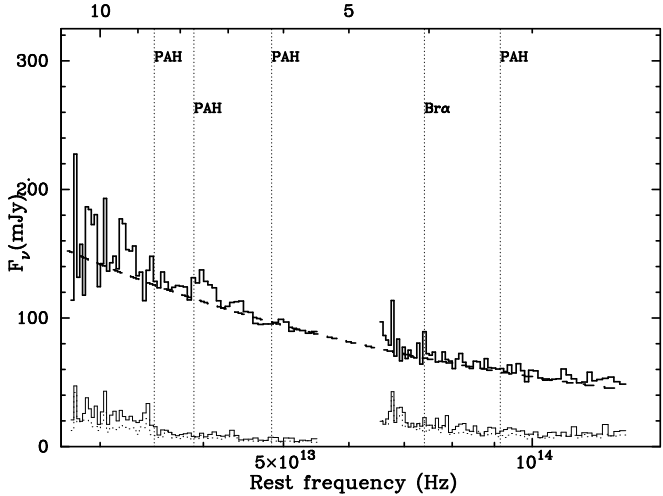}}
      \caption{The weighted averaged mid-IR spectrum of Mrk 279 (dark line),
      the rms deviations about the mean (light line) and the error
      (dotted line). The best-fit power law with index $\alpha=-0.8$
      is shown as a dashed line. The expected positions of PAH and
      \bra\ features are indicated. The x-axis shows the rest frequency
      at the bottom and the rest wavelength, in microns, at the top, both
      represented on a logarithmic scale}
      \label{mean}
   \end{figure}

Figure~\ref{mean} shows the weighted mean 2.5 -- 11.7\,\mic\ spectrum 
of Mrk~279 obtained by averaging all 16 PHT--S spectra.

The spectrum of Mrk~279 is quite similar to the mean Seyfert 1
spectrum obtained by Clavel et al.\ (2000) from their sample of 28
type 1 AGNs. It shows a strong continuum,
with a flux density per frequency unit that drops sharply
with increasing frequency and only weak broad emission features.
The continuum is well approximated by a power law
($F_\nu \propto \nu^{\alpha}$) of spectral index
$\alpha = -0.80 \pm 0.05$ (Fig.~\ref{mean}), close to the average
Seyfert 1 mid-IR
index $\alpha = -0.84 \pm 0.24$ (Clavel et al. (2000)). Its flux
at a fiducial wavelength of 7\,$\mu$ is 103\,mJy. 
While the broad emission features of Polycyclic Aromatic Hydrocarbon (PAH) 
bands (Puget et al.\ 1985) at  3.3\,\mic, 6.2\,\mic, 7.7\,\mic, and 
8.6\,\mic\ are ubiquitous in many different galactic and extragalactic 
line of sight, only the strongest band at 7.7\,\mic\ is unambiguously 
detected in Mrk~279, with an intensity of $0.76\pm0.12$ mJy. Clavel et al.
(2000) showed that PAH emission in AGNs originates in the interstellar 
medium (ISM) of the galaxy, whereas the mid-IR power-law continuum arises 
from near nuclear dust emission in the torus. Because Mrk~279 is a luminous 
AGN, almost a quasar, the apparent weakness of its PAH emission can be 
understood as a contrast effect whereby a faint ISM is observed against 
a bright nucleus.
The 9.7\,\mic\ silicate absorption feature, conspicuous in the mid-IR spectra
of starburst galaxies (Moorwood et al.\ 1996; Rigopoulou et
al.\ 1996; Acosta-Pulido et al.\ 1996), is absent from the Mrk~279
spectrum.




\section{The spectral energy distribution}
\label{sedsec}

\begin{figure}
      \resizebox{0.85\hsize}{!}{\includegraphics{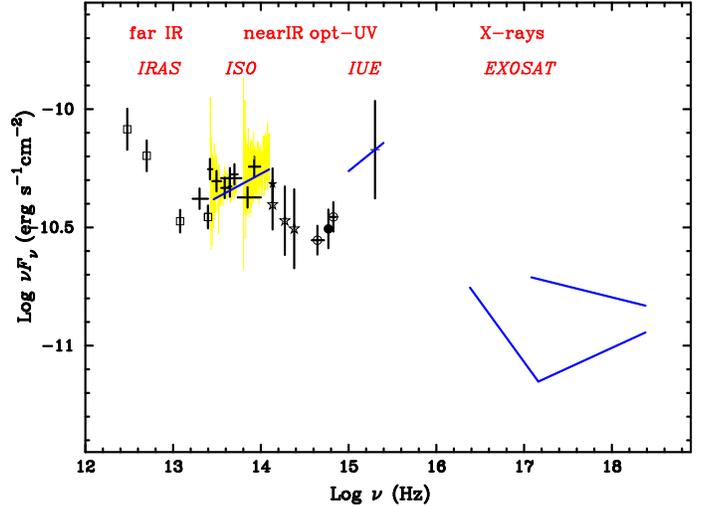}}
      \caption{The spectral energy distribution (SED) of Mrk~279, from
      the far-IR to the X rays. The IRAS data are represented as
      open squares whereas near-IR ground-based data are shown as stars.
      The PHT-S spectrum is shown as a dotted line, while ISOCAM fluxes are
      plotted as crosses where the horizontal bars indicate the filter
      range. The two open circles in the optical are the {\it nuclear\/}
      $R$- and $B$- band fluxes. The filled circle is the mean 5100\,\AA\/
      flux. Optical and near IR data have been corrected for stellar light
      while UV and optical data have been corrected for galactic redenning.
      The best fit power laws mid-IR and UV ($\alpha = -0.7$)
      continua are also displayed. The error bar on the UV flux
      represents the rms fluctuation about the mean 1500\AA\/ due to
      variability. This SED was assembled from data gathered over a
      time span of $\sim$19 years. }
      \label{sed}
   \end{figure}

Figure~\ref{sed} shows the continuum spectral energy distribution (SED)
of Mrk~279 from the far-IR to the X-rays.
The IRAS data points are the average of 6 pointed observations reported
by Edelson \& Malkan (1987). The near-IR data are from Spinoglio et al.
(1985) after subtraction of the stellar light (\S{\ref{irgal}}). The 
$R$ and $B$ band fluxes (Granato et al.\ 1993)
and the mean 5100\,\AA-flux (this paper) have also been corrected for
the underlying galaxy contribution and de-redenned. The 1500~\AA\ data point
represents the average of 26 observations made with IUE (Rodriguez et al. 
1998) between 1978 and 1991. It has been corrected for foreground redenning
using $N_{\rm H} = 1.6\times10^{20}$\,cm$^{-2}$ (Elvis et al.\ 1989). The 
large error bar reflects the strong variability of Mrk~279 at UV wavelengths.
The EXOSAT X-ray data are from 1983 and 1984 and are best described
in terms of a  broken power-law (Ghosh \& Soundararajaperumal 1992),
while the 1994 data from ASCA are modeled with a unique power law
in the Tartarus Database ({\sf http://tartarus.gsfc.nasa.gov/}).

Since this SED is constructed from data collected over $\sim$~19
years, we caution that it may be distorted by variability.
Variability is important in shaping the X-ray, ultraviolet, and optical
spectrum, but is much less significant at longer wavelengths 
(see \S \ref{irvarsec}).
Bearing these limitations in mind, it is still possible
to draw some general conclusions which are not affected by flux variations
at short wavelengths.

The Mrk~279 SED displays three broad maxima or
``bumps''.
The first maximum occurs in the far-IR at wavelengths $\geq\,25\,\mic$.
Given the large IRAS ($\sim1'$) apertures and
the cold color temperature of the far-IR bump,
the 100\,\mic\ and 60\,\mic\ fluxes are probably dominated
by cold dust from the host galaxy's ISM.
Hence, the far-IR ``bump'' is most likely not related to the AGN itself.
The second maximum is the usual ``big blue bump'' which dominates
the SED of type 1 AGNs from the optical to the soft X-rays.
It is usually identified as thermal emission from an accretion disk.
In between these two maxima lies a third and smaller bump
which extends from $\sim1$\,\mic\ to $\sim15$--20\,\mic.
We tentatively identify this mid-IR bump as
thermal emission from dust in
the putative molecular torus and/or from dust in the NLR,
as discussed below.

\section{The IR variability}
\label{irvarsec}

The total duration of our ISO campaign was 374 days,
with a mean sampling interval of 24.9 days.
In addition to the weighted-mean spectrum, $\langle F_\nu \rangle$,
Fig.~\ref{mean} displays the
rms spectrum  $\langle {\rm rms}_\nu \rangle$,
and the mean error spectrum $\epsilon_\nu$.
The latter was evaluated as
\begin{equation}
\epsilon^2_\nu = \frac{1}{N\,\sum_{i=1}^{N} 1/\Delta_{\nu,i}^2}
\end{equation}
where $\Delta_{\nu,i}$ is the uncertainty associated with the flux $F_i$ at
epoch $i$ and $N\,=\,16$ is the number of epochs. We can exclude significant 
(i.e. 3$\sigma$) flux variability since the largest value of 
$\langle {\rm rms}_\nu \rangle / \epsilon_\nu$ is only $\approx $ 1.4. 
A chi-square test was also applied to the average PHT-S fluxes of
Table~\ref{lcur}. The reduced chi-squares are $<~1$ which confirms 
that no significant variations of the mid-IR flux 
took place during our $\sim1$-year observing campaign.


\section{Optical variability and the H$\beta$ emission-line lag}

\begin{figure}
      \resizebox{\hsize}{!}{\includegraphics{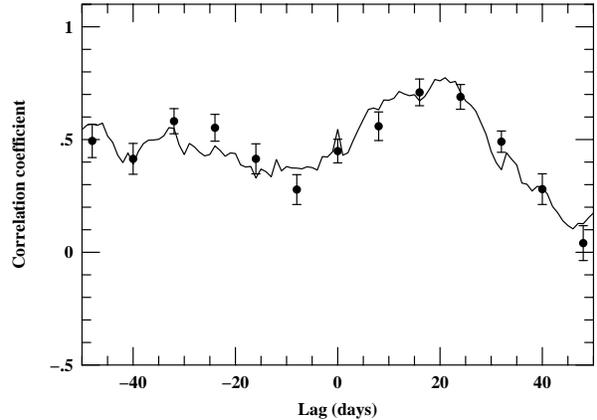}}
      \caption{Cross-correlation function of the \hb\ line intensity
      versus the 5100~\AA\ optical continuum flux between JD 2,450,095.1
      and JD 2,450,289.7 (Fig.~\ref{lcplot}). The interpolation CCF is
      shown as a solid line and the points with error bars are
      computed using the DCF method (with a bin size of 8 days).
      The \hb\ line lags behind the continuum by about 16 days.}
      \label{ccfplot}
   \end{figure}

The relatively dense sampling of the optical light curves between 
January and July 1996 allows us to
measure the time-delayed response, or ``lag'', of
the \hb\ emission line to continuum variations by
cross-correlation of the light curves shown in Fig.~\ref{lcplot}.
We used both  the interpolation method of Gaskell \& Sparke (1986)
and the discrete-correlation function (DCF) method
of Edelson \& Krolik (1988), in both cases
employing the specific implementation described by
White \& Peterson (1994).
The centroid of the cross-correlation is at $16.7^{+5.3}_{-5.6}$\,days.
Uncertainties were estimated using the model-independent
FR/RSS Monte-Carlo method described by Peterson et al.\ (1998).
The cross-correlation results are shown in
Table~\ref{crossc} and Fig.~\ref{ccfplot}. For comparison, 
Table~\ref{crossc} also lists the results of previous Mrk~279 monitoring 
campaigns by Maoz et al.\ (1990) and Stirpe et al.\ (1994).

\begin{center}
\begin{table*}[htbp]
\caption{Cross-Correlation Results}
\label{crossc}
\begin{tabular}{lcccc}
\multicolumn{1}{c}{ } &
\multicolumn{1}{c}{This Work (\hb)} &
\multicolumn{1}{c}{Wise Obs.(\hb)} &
\multicolumn{1}{c}{Wise Obs.(\ha)} &
\multicolumn{1}{c}{LAG (\ha)}  \\
\multicolumn{1}{c}{Parameter} &
\multicolumn{1}{c}{(1996)} &
\multicolumn{1}{c}{(1988)} &
\multicolumn{1}{c}{(1988)} &
\multicolumn{1}{c}{(1990)} \\
\multicolumn{1}{c}{(1)} &
\multicolumn{1}{c}{(2)} &
\multicolumn{1}{c}{(3)} &
\multicolumn{1}{c}{(4)} &
\multicolumn{1}{c}{(5)} \\
Cross-correlation centroid $\tau_{\rm cent}$ (days)&$16.7^{+5.3}_{-5.6}$
                                                &$2.5^{+25.5}_{-5.4}$
                                                &$11.6^{+8.5}_{-11.7}$
                                                &$6.8^{+19.8}_{-6.9}$ \\
Cross-correlation peak $\tau_{\rm peak}$ (days) &$21^{+2}_{-9}$
                                                &$3^{+28}_{-5}$
                                                &$11^{+11}_{-11}$
                                                &$2^{+29}_{-3}$ \\
Peak correlation coefficient $r_{\rm max}$      &0.769
                                                &0.799
                                                &0.793
                                                &0.795 \\
Continuum rms fractional variability $F_{\rm var}$ &0.093
                                                &0.071
                                                &0.071
                                                &0.126 \\
Mean time between observations (days)           &2.3
                                                &4.1
                                                &4.1
                                                &6.0 \\
Duration of campaign (days)                     & 195
                                                & 156
                                                & 156
                                                & 152 \\
\end{tabular}
\end{table*}
\end{center}

\section{Discussion and concluding remarks}

The mid-IR spectrum of Mrk~279 
shows a strong power-law continuum
of spectral index $-0.80\pm0.05$, with weak PAH emission
bands and no detectable silicate $\sim 9.7\,\mic$ feature.
The mid-IR bump of Mrk~279 extends\footnote{for the extension of 
the bump we consider the wavelength range $\Delta\lambda$ over which
$\nu F_\nu$ exceeds one third of its peak value.}
from roughly 1.25\,\mic\ to 15--20\,\mic\ and is 
wider than a single blackbody. It peaks near $\approx 3\,\mic$ .
In Fairall~9, the mid-IR bump most likely originates from the
re-processing of UV and optical photons by nuclear dust (\S{1}).
We can estimate the distance to the central source
of the innermost and hottest dust grains in Mrk~279,
$r_{\rm in}$, by scaling directly from Fairall~9
(Clavel et al.\ 1989). Mrk~279 is
approximately eight times less luminous than Fairall~9.
Since the inner radius of the dust distribution is presumably
controlled by sublimation,
$r_{\rm in}$ should scale approximately as $L^{1/2}$, so
$r_{\rm in}$ should be a factor of $\simeq2.8$ smaller in Mrk~279
than in Fairall~9, i.e.
$r_{\rm in} \approx 140\pm36$~light-days.

During the ISO campaign, the mid-IR flux did not experience
variations of amplitude larger than 10\,\%, the detection limit of
the PHT-S instrument.
Optical data contemporaneous to the IR observations revealed
significant fluctuations of the 5100\,\AA\ flux with a relative
rms amplitude of 9\% and a ratio of the maximum
to the minimum fluxes, R$_{\rm max}\,\sim\,1.64\pm0.11$.
Any upper limit to the mid-IR variability in Mrk~279 has to
be examined in the light of the UV and optical continuum variations
over the same period of time. As noted earlier, in the dust reprocessing
scenario the amplitude of the MIR flux variations will be reduced compared 
to that of the primary UV-optical source because of the finite propagation 
time of the photons. Imagine a short (duration $\leq 1$ day) pulse
of the UV-optical source illuminating a thin dust annulus,
inclined by $i = 10^{\circ}$ with respect to the line of sight.
The annulus IR response will be delayed by
$\delta t\,=\,(1-\sin{i}) r_{\rm in}/c$
and will last for $2(\sin{i}) r_{\rm in}/c$.
For numerical values appropriate to Mrk~279,
the duration of the IR reverberated pulse will be 38 days
and its peak amplitude thereby reduced by a factor of order 38.
Given the relatively low amplitude of the optical flux variations
(Fig.~\ref{lcplot}), the absence of measurable variations of
the MIR flux is consistent with the above scenario.

Though a detailed quantitative fit with a particular model is
beyond the scope of this paper, it is nevertheless illustrative
to perform a qualitative comparison of our data
with the theoretical predictions from the torus model
by Pier \& Krolik (1992). This model predicts a mid-IR ``bump''
that is approximately 0.7 to 1 decade wide in wavelengths, in
agreement with the Mrk~279 observations. In the Pier \& Krolik (1992) model,
the torus emission is expected to peak at a wavelength $\lambda_{\rm peak}$
that depends primarily on the flux illuminating the torus inner surface
and its inclination angle $i$ with respect to the line of sight.
The relatively high color temperature implied by
$\lambda_{\rm peak} \approx3\,\mic$ constrains the inclination
to be small ($\cos{i} \geq 0.75$). The absence of silicate absorption
also rules-out very optically thick models and constrains
the vertical column density at the torus inner edge, 
${\rm N_{H}\,\leq\,10^{24}\,cm^{-2}}$.
Comparison of Fig.~\ref{sed} with Fig.~5 of Pier \& Krolik (1992)
also suggests a moderately thick torus, with $r_{\rm in}/h = 0.3$.

The delay $\Delta$T of \hb\ w.r.t. the optical continuum
was $16.7^{+5.3}_{-5.6}$ days during this campaign.
Comparison with the results from previous monitoring campaigns
(see Table~\ref{crossc}) does not reveal any significant change
of $\Delta$T over a time span of $\sim$8 years. In other words,
we find no evidence for a secular change in the structure of the
BLR in Mrk~279. Equating ${\rm c\,\times\,\Delta T}$ with the
emissivity weighted radius  ${\rm R_{BLR}}$ of the \hb\ emitting
region, one sees that $r_{\rm in}$ is about 8 times larger than
${\rm R_{BLR}}$. In other words, the BLR lies well within the
dust evaporation radius.

\begin{acknowledgements}
The authors are grateful to all the observatories involved for the generous
allocation of observing time
and Jos\'e Acosta-Pulido for helpful
discussions on the PHT-S instrument calibration.
MS acknowledges partial support by Spanish CICYT grant PB-ESP95-0389-C02-02
and all the staff at the {\it Laboratorio de Astrof\'{\i}sica Espacial y
  F\'{\i}sica Fundamental, Spain} where most of this work was done.
Support for the ground-based observations was provided by
the National Science Foundation through grant AST--9420080 to
Ohio State University. Observations at the Wise Observatory are supported
by grants from the Israel Science Foundation. This research has made use of
the TARTARUS database, which is supported by Jane Turner and Kirpal
Nandra under NASA grants NAG5-7385 and NAG5-7067.
\end{acknowledgements}

\newpage
{\bf Electronic tables}

\begin{table*}[htbp]
\caption{Optical Photometry from Wise Observatory}
\label{photom}
\begin{tabular}{lccccc}
\multicolumn{1}{c}{UT} &
\multicolumn{1}{c}{Julian Date} &
\multicolumn{1}{c}{$B^{\star}$} &
\multicolumn{1}{c}{$V^{\star}$} &
\multicolumn{1}{c}{$R^{\star}$} &
\multicolumn{1}{c}{$I^{\star}$} \\
\multicolumn{1}{c}{Date} &
\multicolumn{1}{c}{($-2,450,000$)} &
\multicolumn{1}{c}{(mag)} &
\multicolumn{1}{c}{(mag)} &
\multicolumn{1}{c}{(mag)} &
\multicolumn{1}{c}{(mag)} \\
\multicolumn{1}{c}{(1)} &
\multicolumn{1}{c}{(2)} &
\multicolumn{1}{c}{(3)} &
\multicolumn{1}{c}{(4)} &
\multicolumn{1}{c}{(5)} &
\multicolumn{1}{c}{(6)} \\
1995 Dec 28& 80.6&$13.838\pm0.020$&$12.533\pm0.016$&$12.461\pm0.012$& \\
1996 Jan 29&112.4&$14.111\pm0.040$&$12.702\pm0.018$&$12.578\pm0.010$&
                    $12.771\pm0.022$ \\
1996 Mar 27&170.6&$14.097\pm0.015$&$12.706\pm0.022$&$12.588\pm0.016$& \\
1996 Mar 28&171.5&$14.070\pm0.021$&$12.715\pm0.022$&$12.605\pm0.012$& \\
1996 Mar 29&172.5&$14.151\pm0.013$&$12.805\pm0.005$&$12.699\pm0.006$& \\
1996 Mar 31&174.5&$14.110\pm0.024$&$12.738\pm0.021$&$12.631\pm0.023$&
                 $12.829\pm0.023$ \\
1996 May  2&206.5&$14.517\pm0.015$&$13.243\pm0.014$&$13.130\pm0.012$&
                   $13.426\pm0.026$ \\
1996 Jul  1&266.3&$14.377\pm0.042$&$12.697\pm0.056$&$12.733\pm0.098$&  \\
1996 Jul  2&267.3&$14.417\pm0.030$&                &$12.762\pm0.051$&
                   $12.964\pm0.080$ \\
1996 Aug  6&302.2&$14.111\pm0.017$&$12.745\pm0.014$&$12.626\pm0.017$&
                   $12.831\pm0.018$ \\
1996 Dec 19&437.6&$14.245\pm0.045$&$12.810\pm0.037$&$12.659\pm0.022$&
                   $12.837\pm0.012$ \\
1997 Feb 18&498.6&$14.192\pm0.013$&$12.787\pm0.008$&$12.673\pm0.019$&
                   $12.856\pm0.024$ \\
1997 Mar 19&527.5&$14.179\pm0.019$&$12.769\pm0.014$&$12.661\pm0.009$&
                   $12.842\pm0.021$ \\
1997 May 10&579.4&$14.090\pm0.027$&$12.684\pm0.014$&$12.576\pm0.009$&
                   $12.776\pm0.013$ \\
1997 Jun  9&609.4&$14.088\pm0.021$&$12.701\pm0.013$&$12.604\pm0.012$& \\
1997 Jul 25&655.3&$14.005\pm0.028$&$12.652\pm0.016$&$12.550\pm0.021$
                   &$12.770\pm0.013$  \\
1997 Aug 30&691.3&$14.118\pm0.030$&$12.705\pm0.010$&$12.578\pm0.020$&  \\
\multicolumn{6}{l}{$\star$ Instrumental magnitudes, not scaled to any
standard system} \\
\end{tabular}
\end{table*}

\newpage
\begin{center}
\begin{table}[htp]
\caption{Optical Continuum and \hb\ fluxes}
\label{optlc}
\begin{tabular}{ccc}
\multicolumn{1}{c}{Julian Date} &
\multicolumn{1}{c}{$F_{\lambda}$\,(5100\,\AA)} &
\multicolumn{1}{c}{$F(\hb)$} \\
\multicolumn{1}{c}{($-2,450,000$)} &
\multicolumn{1}{c}{(a)} &
\multicolumn{1}{c}{(b)} \\
\multicolumn{1}{c}{(1)} &
\multicolumn{1}{c}{(2)} &
\multicolumn{1}{c}{(3)} \\
  95.1& $  9.39 \pm    0.40$&   $  6.56 \pm    0.28$ \\
  96.0& $  9.67 \pm    0.41$&   $  6.57 \pm    0.28$ \\
  97.0& $  8.59 \pm    0.37$&   $  6.10 \pm    0.26$ \\
  98.1& $  9.09 \pm    0.39$&   $  6.30 \pm    0.27$ \\
 104.0& $  7.96 \pm    0.35$&   $  6.36 \pm    0.28$ \\
 108.0& $  7.84 \pm    0.35$&   $  6.32 \pm    0.27$ \\
 109.0& $  7.37 \pm    0.33$&   $  6.18 \pm    0.27$ \\
 110.0& $  7.74 \pm    0.35$&   $  6.25 \pm    0.27$ \\
 110.4& $  7.29 \pm    0.22$&   $  6.58 \pm    0.20$ \\
 111.0& $  7.57 \pm    0.34$&   $  6.38 \pm    0.28$ \\
 112.1& $  7.54 \pm    0.34$&   $  6.07 \pm    0.26$ \\
 119.0& $  7.20 \pm    0.14$&   $  6.66 \pm    0.13$ \\
 123.0& $  6.99 \pm    0.14$&   $  6.26 \pm    0.12$ \\
 124.0& $  7.08 \pm    0.32$&   $  6.19 \pm    0.27$ \\
 128.0& $  7.08 \pm    0.32$&   $  5.54 \pm    0.24$ \\
 129.0& $  7.46 \pm    0.34$&   $  5.86 \pm    0.25$ \\
 130.0& $  7.12 \pm    0.33$&   $  5.63 \pm    0.24$ \\
 131.0& $  7.53 \pm    0.34$&   $  5.90 \pm    0.26$ \\
 132.0& $  7.51 \pm    0.34$&   $  5.69 \pm    0.25$ \\
 134.0& $  7.18 \pm    0.33$&   $  5.55 \pm    0.24$ \\
 134.9& $  7.25 \pm    0.33$&   $  5.75 \pm    0.25$ \\
 136.9& $  7.42 \pm    0.15$&   $  6.07 \pm    0.12$ \\
 143.0& $  6.99 \pm    0.14$&   $  5.62 \pm    0.11$ \\
 149.9& $  6.80 \pm    0.14$&   $  6.05 \pm    0.12$ \\
 156.5& $  7.11 \pm    0.21$&   $  6.02 \pm    0.18$ \\
 164.9& $  7.11 \pm    0.14$&   $  5.90 \pm    0.12$ \\
 168.9& $  7.34 \pm    0.14$&   $  5.83 \pm    0.11$ \\
 170.0& $  7.42 \pm    0.34$&   $  5.58 \pm    0.24$ \\
 171.0& $  7.51 \pm    0.34$&   $  5.88 \pm    0.25$ \\
 175.9& $  7.27 \pm    0.14$&   $  6.07 \pm    0.12$ \\
 185.9& $  6.73 \pm    0.12$&   $  6.24 \pm    0.11$ \\
 188.0& $  6.51 \pm    0.31$&   $  5.79 \pm    0.25$ \\
 189.0& $  6.07 \pm    0.29$&   $  5.47 \pm    0.24$ \\
 190.0& $  6.46 \pm    0.31$&   $  5.94 \pm    0.26$ \\
 191.0& $  6.61 \pm    0.31$&   $  6.12 \pm    0.26$ \\
 191.9& $  6.31 \pm    0.12$&   $  5.96 \pm    0.11$ \\
 193.0& $  6.62 \pm    0.31$&   $  5.95 \pm    0.26$ \\
 194.0& $  6.11 \pm    0.29$&   $  6.09 \pm    0.26$ \\
 198.8& $  6.07 \pm    0.11$&   $  5.82 \pm    0.11$ \\
 199.8& $  6.53 \pm    0.31$&   $  5.97 \pm    0.26$ \\
 206.8& $  6.48 \pm    0.13$&   $  5.47 \pm    0.11$ \\
 211.8& $  6.55 \pm    0.31$&   $  5.70 \pm    0.25$ \\
 212.9& $  6.41 \pm    0.12$&   $  5.47 \pm    0.10$ \\
 213.9& $  6.79 \pm    0.32$&   $  5.60 \pm    0.24$ \\
 215.4& $  6.40 \pm    0.19$&   $  5.26 \pm    0.16$ \\
 218.6& $  7.16 \pm    0.33$&   $  5.24 \pm    0.23$ \\
 219.6& $  7.11 \pm    0.33$&   $  5.77 \pm    0.25$ \\
 220.9& $  7.06 \pm    0.14$&   $  5.33 \pm    0.11$ \\
 221.6& $  6.95 \pm    0.32$&   $  5.09 \pm    0.22$ \\
 222.7& $  6.95 \pm    0.32$&   $  5.26 \pm    0.23$ \\
 223.6& $  7.26 \pm    0.33$&   $  5.49 \pm    0.24$ \\
 224.6& $  6.95 \pm    0.32$&   $  5.44 \pm    0.24$ \\
\\
\multicolumn{3}{l}{(a) \funits{15}} \\
\multicolumn{3}{l}{(b) \lunits{13}} \\
\end{tabular}
\end{table}
\end{center}


\begin{center}
\addtocounter{table}{-1}
\begin{table}[htp]
\caption{Optical Continuum  and \hb\ fluxes (cont.)}
\begin{tabular}{ccc}
\multicolumn{1}{c}{Julian Date} &
\multicolumn{1}{c}{$F_{\lambda}$\,(5100\,\AA)} &
\multicolumn{1}{c}{$F(\hb)$} \\
\multicolumn{1}{c}{($-2,450,000$)} &
\multicolumn{1}{c}{(a)} &
\multicolumn{1}{c}{(b)} \\
\multicolumn{1}{c}{(1)} &
\multicolumn{1}{c}{(2)} &
\multicolumn{1}{c}{(3)} \\
 225.6& $  6.88 \pm    0.32$&   $  5.33 \pm    0.23$ \\
 226.6& $  6.87 \pm    0.32$&   $  5.24 \pm    0.23$ \\
 227.8& $  6.84 \pm    0.14$&   $  5.43 \pm    0.11$ \\
 231.6& $  7.06 \pm    0.32$&   $  5.69 \pm    0.25$ \\
 233.8& $  6.82 \pm    0.14$&   $  5.57 \pm    0.11$ \\
 238.3& $  6.51 \pm    0.19$&   $  5.82 \pm    0.17$ \\
 240.6& $  6.70 \pm    0.31$&   $  6.04 \pm    0.26$ \\
 241.8& $  6.67 \pm    0.13$&   $  5.69 \pm    0.11$ \\
 242.6& $  6.82 \pm    0.32$&   $  6.20 \pm    0.27$ \\
 248.8& $  6.71 \pm    0.13$&   $  5.47 \pm    0.11$ \\
 249.7& $  6.30 \pm    0.30$&   $  5.84 \pm    0.25$ \\
 250.7& $  6.23 \pm    0.30$&   $  5.44 \pm    0.24$ \\
 251.7& $  6.84 \pm    0.32$&   $  5.92 \pm    0.26$ \\
 253.6& $  5.85 \pm    0.28$&   $  5.63 \pm    0.24$ \\
 253.8& $  6.01 \pm    0.12$&   $  5.55 \pm    0.11$ \\
 254.6& $  5.90 \pm    0.29$&   $  5.84 \pm    0.25$ \\
 256.6& $  6.43 \pm    0.30$&   $  5.64 \pm    0.25$ \\
 257.7& $  6.63 \pm    0.31$&   $  5.80 \pm    0.25$ \\
 258.6& $  6.24 \pm    0.30$&   $  5.30 \pm    0.23$ \\
 259.6& $  6.04 \pm    0.29$&   $  5.25 \pm    0.23$ \\
 262.8& $  6.44 \pm    0.13$&   $  5.35 \pm    0.11$ \\
 270.7& $  6.20 \pm    0.30$&   $  5.27 \pm    0.23$ \\
 272.3& $  6.83 \pm    0.20$&   $  5.07 \pm    0.15$ \\
 280.6& $  6.62 \pm    0.31$&   $  5.13 \pm    0.22$ \\
 281.7& $  6.80 \pm    0.32$&   $  5.53 \pm    0.24$ \\
 282.7& $  6.05 \pm    0.29$&   $  5.01 \pm    0.22$ \\
 283.7& $  6.07 \pm    0.29$&   $  5.19 \pm    0.22$ \\
 284.7& $  6.51 \pm    0.31$&   $  5.36 \pm    0.23$ \\
 286.7& $  6.86 \pm    0.32$&   $  5.28 \pm    0.23$ \\
 287.6& $  6.54 \pm    0.31$&   $  4.96 \pm    0.22$ \\
 288.7& $  6.94 \pm    0.32$&   $  5.30 \pm    0.23$ \\
 289.7& $  6.46 \pm    0.31$&   $  5.04 \pm    0.22$ \\
 460.1& $  6.46 \pm    0.31$&   $  5.54 \pm    0.24$ \\
 465.0& $  6.44 \pm    0.30$&   $  5.55 \pm    0.24$ \\
 480.1& $  5.98 \pm    0.29$&   $  5.66 \pm    0.25$ \\
 481.1& $  6.39 \pm    0.30$&   $  5.46 \pm    0.24$ \\
 482.1& $  6.36 \pm    0.30$&   $  5.75 \pm    0.25$ \\
 485.9& $  6.51 \pm    0.31$&   $  5.84 \pm    0.25$ \\
 487.0& $  6.43 \pm    0.30$&   $  5.94 \pm    0.26$ \\
 489.0& $  6.58 \pm    0.31$&   $  5.79 \pm    0.25$ \\
 489.9& $  6.79 \pm    0.32$&   $  5.96 \pm    0.26$ \\
 491.0& $  6.55 \pm    0.31$&   $  5.49 \pm    0.24$ \\
 494.0& $  6.30 \pm    0.30$&   $  5.42 \pm    0.23$ \\
 510.0& $  6.66 \pm    0.31$&   $  5.60 \pm    0.24$ \\
 519.8& $  6.61 \pm    0.31$&   $  5.30 \pm    0.23$ \\
 546.4& $  7.42 \pm    0.22$&   $  4.88 \pm    0.15$ \\
 547.8& $  7.61 \pm    0.34$&   $  5.14 \pm    0.22$ \\
 548.8& $  7.43 \pm    0.34$&   $  5.21 \pm    0.23$ \\
 549.8& $  7.64 \pm    0.34$&   $  5.39 \pm    0.23$ \\
 550.8& $  7.64 \pm    0.34$&   $  5.59 \pm    0.24$ \\
 551.8& $  7.15 \pm    0.33$&   $  5.24 \pm    0.23$ \\
 552.8& $  7.49 \pm    0.34$&   $  5.55 \pm    0.24$ \\
 569.8& $  7.55 \pm    0.34$&   $  5.43 \pm    0.23$ \\
\\
\multicolumn{3}{l}{(a) \funits{15}} \\
\multicolumn{3}{l}{(b) \lunits{13}} \\
\end{tabular}
\end{table}
\end{center}

\begin{center}
\addtocounter{table}{-1}
\begin{table}[htp]
\caption{Optical Continuum  and \hb\ fluxes (cont.)}
\begin{tabular}{ccc}
\multicolumn{1}{c}{Julian Date} &
\multicolumn{1}{c}{$F_{\lambda}$\,(5100\,\AA)} &
\multicolumn{1}{c}{$F(\hb)$} \\
\multicolumn{1}{c}{($-2,450,000$)} &
\multicolumn{1}{c}{(a)} &
\multicolumn{1}{c}{(b)} \\
\multicolumn{1}{c}{(1)} &
\multicolumn{1}{c}{(2)} &
\multicolumn{1}{c}{(3)} \\
 575.4& $  7.78 \pm    0.23$&   $  5.85 \pm    0.17$ \\
 578.8& $  7.83 \pm    0.35$&   $  5.37 \pm    0.23$ \\
 580.5& $  7.39 \pm    0.22$&   $  5.46 \pm    0.16$ \\
 581.8& $  7.22 \pm    0.33$&   $  5.52 \pm    0.24$ \\
 582.6& $  6.96 \pm    0.32$&   $  5.57 \pm    0.24$ \\
 597.7& $  7.08 \pm    0.32$&   $  5.56 \pm    0.24$ \\
 602.6& $  6.36 \pm    0.30$&   $  5.00 \pm    0.22$ \\
 604.6& $  6.68 \pm    0.31$&   $  5.22 \pm    0.23$ \\
 605.7& $  7.11 \pm    0.33$&   $  4.93 \pm    0.21$ \\
 608.7& $  7.41 \pm    0.34$&   $  5.09 \pm    0.22$ \\
 609.6& $  7.34 \pm    0.33$&   $  5.35 \pm    0.23$ \\
 610.6& $  6.88 \pm    0.32$&   $  4.99 \pm    0.22$ \\
 627.6& $  7.45 \pm    0.34$&   $  5.35 \pm    0.23$ \\
 628.6& $  7.58 \pm    0.34$&   $  5.51 \pm    0.24$ \\
 629.7& $  8.26 \pm    0.36$&   $  5.83 \pm    0.25$ \\
 630.7& $  7.92 \pm    0.35$&   $  5.98 \pm    0.26$ \\
\\
\multicolumn{3}{l}{(a) \funits{15}} \\
\multicolumn{3}{l}{(b) \lunits{13}} \\
\end{tabular}
\end{table}
\end{center}

\end{document}